# Enhancement of photocatalytic performance of $V_2O_5$ by rare-earth ions doping, synthesized by facile hydrothermal technique

Mohammad Humaun Kabir[1,2], M. Z. Hossain[1], M. A. Jalil[1], M.M. Hossain[1], M. A. Ali[1], M. U. Khandaker[3], D. Jana[4], Md. Motinur Rahman[5], M. Khalid Hossain[5], M.M. Uddin[1*]

[1]Department of Physics, Chittagong University of Engineering and Technology (CUET), Chattogram 4349, Bangladesh

[2]Department of Materials Science & Engineering, Chittagong University of Engineering and Technology (CUET), Chattogram 4349, Bangladesh

[3]Centre for Applied Physics and Radiation Technologies, School of Engineering and Technology, Sunway University, 47500 Bandar Sunway, Selangor, Malaysia

[4]Department of Physics, University of Calcutta, 92 A P C Road, Kolkata-700009, West Bengal, India

[5]Institute of Electronics, Atomic Energy Research Establishment, Bangladesh Atomic Energy Commission, Dhaka 1349, Bangladesh

E-mail: mohi@cuet.ac.bd (Md. Mohi Uddin)

**Abstract:**

The rare-earth (RE) elements [Holmium (Ho) and Ytterbium (Yb)] doped vanadium pentoxide ($V_2O_5$) with a series of doping concentrations (1 mol.%, 3 mol.%, and 5 mol.%) have been successfully synthesized using environment-friendly facile hydrothermal method. The effect of RE ions on the photocatalytic efficiency of doped $V_2O_5$ has also been analyzed. The stable orthorhombic crystal structure of doped $V_2O_5$ confirms by the X-ray diffraction with no secondary phase, and high-stressed conditions are generated for the 3 mol.%. The crystallite size, strain, and dislocation density are calculated to perceive the doping effect on the bare $V_2O_5$. The optical characteristics have been measured using UV-vis spectroscopy. The absorptions are found to be increased with increasing doping concentrations; however, the bandgap remains in the visible range. The photocatalytic properties are examined for the compounds with varying pH, and it is observed that higher efficiency is exhibited for the pH 7 and catalyst concentration 500 ppm. The highest degradation efficiency is found to be 93% and 95% for the 3 mol.% of Ho and Yb-doped $V_2O_5$ samples within 2 hours, respectively. It is elucidated that the RE ions significantly impact the catalytic behavior of $V_2O_5$, and the mechanism behind these extraordinary efficiencies has been explained thoroughly.

## 1. Introduction

One of the most insistent and concerning issues on the earth is water pollution, which has been rising sharply day by day. The textile, pesticide, tanning, and dye industries are the leading polluters of surface water. Unfortunately, natural fighters, such as bacteria or viruses cannot heal this chemically-tainted water .[1]However, one of the effective ways to remedy this problem is photocatalytic materials, which can generate free radicals and decompose these pollutants more easily with the aid of solar radiation at ambient temperature.[2,3]The nanoscale metal oxide and their nanocomposites have been considered as the best candidates for photocatalysis due to their extended surface area and recyclability.[4] Some commonly used photocatalysts are $TiO_2$, ZnO, $Fe_2O_3$, $SnO_2$, $VO_x$, $MoO_x$, etc.[5,6]

The vanadium pentoxide ($V_2O_5$), a compound of the $VO_x$ family, has been an emerging applicant in recent years in the field of lithium-ion batteries, solar cells, gas sensors, catalysts, and optoelectronics due to its high physiochemical stability, large absorption efficiency in the solar spectrum, and non-toxic nature along with biocompatibility.[7–10] It has an orthorhombic structure with a direct band gap of 2.2-2.8 eV, which makes it suitable for photocatalysis.[11,12] Besides, inherent oxygen vacancies of $V_2O_5$ lead to the form of many planes of defects to arise between the valence band and conduction band, accordingly creating a quick recombination pathway and aiding the photocatalytic process.[13,14] However, recent findings of $V_2O_5$ indicate that photo-generated electrons and holes quickly recombine and eventually limit the photodegradation efficiency. To achieve a highly efficient photocatalytic reaction, it is utterly necessary to improve the numerous electron-hole pairs that boost charge separation and impede charge carrier recombination.[6,15,16]

Previously, cation doping has been used prominently to reduce or resolve the quick recombination problem. Three categories, including the main group (Na[17], Mg[18], Al[19]), transition metals (Ti[20], Mn[21], Fe[22], Co[23], Ni[24], Cu[25], Zn[26], Zr[27], Mo, and Sn[28,29]), and rare-earths (RE)(Y[30], La[31], Ce[32], Nd[33],Ga[34],and Gd[35]) have been documented for cation doping in the $V_2O_5$ nanoparticles. Almost all the reports dealt with the storage capacity and its associated properties. Very few of them were looking for the enhanced photocatalytic behavior of doped $V_2O_5$. As pointed out that the main group metals (like $Al^{3+}$) act as electron sinks in the doped $V_2O_5$, resulting in its inability to improve the photocatalytic properties.[19] On the other hand, transitional metal doping enhances the oxidizing capability and the number of active sites by altering surface morphology and their band structures; consequently, not being involved in the improvement of the photocatalytic activity.[25,35]

The RE elements doping has become a pronounced technique to enhance photocatalytic activity due to its inherent nature of 4f orbitals. Having a partially-filled 4f orbital along with an unfilled 5d orbital, they can easily attack organic dyes and pollutants strongly by forming complex compounds.[16] Moreover, the RE metals doping can also be modified the bandgap of the catalyst and shift to the visible range, which plays a vital role in improving the photocatalytic activities of the catalyst.[31,33] In addition, the RE doping increases the oxygen vacancies in the semiconductor oxides and hence, enhances the newly formed vacancies that can trap the photo-generated electrons, leading to a lower recombination rate of electron-hole pair.[34] Furthermore, Because RE metals have a smaller work function than metal oxides, they are able to pull electrons from the surface of the metal oxides, which speeds up the photocatalytic activity.[16] On account of these points, it is urgent to study the photocatalytic activity of RE ions doped $V_2O_5$ since very few reports have been published yet in this regard.

The photocatalytic properties of $V_2O_5$ can also be tuned by controlling its morphological structure, such as nanorods[35], nanodots[36], nanowires[37], nanoflowers[38], etc. This shape-controlled $V_2O_5$ can be synthesized by different methods, such as sol-gel synthesis, hydrothermal method, chemical vapor deposition, electrochemical deposition, pulsed laser ablation, solution combustion, etc.[39–41] Among them, the hydrothermal method is reported to be a highly productive, low-cost, easy-handling, and environment-friendly synthesis procedure. Moreover, its low-temperature operating system aids in guiding the crystallinity and morphology of the compounds that eventually regulate the photocatalytic activity.[24,37]

Although various metal doping has been implanted into the $V_2O_5$ for enhancing energy storage performance, a few reports have been focused on their photocatalytic performance. Rigorous studies of the photocatalytic performance of $V_2O_5$, especially RE elements doping, are required to suggest $V_2O_5$ as a potential catalyst for pollutants or dye degradation. In this paper, we are introducing RE elements, such as Holmium (Ho) and Ytterbium (Yb) for the first time with different weight ratios (1, 3 & 5 wt.%) into the $V_2O_5$ nanoparticles by following hydrothermal technique to evaluate its photocatalytic efficiency. Here, we also explore the effect of doping on the structural and morphological properties of pristine $V_2O_5$. Moreover, optical properties, like absorbance, reflectance, and optical bandgap, are studied thoroughly to explain its dye degradation performance.

## 2. Experimental Details

### 2.1 Chemicals

Commercially available ammonium metavanadate ($NH_4VO_3$), holmium nitrate pentahydrate ($Ho(NO_3)_3 \cdot 5H_2O$) and ytterbium nitrate pentahydrate ($Yb(NO_3)_3 \cdot 5H_2O$) were used without any further purification. All the chemicals are purchased from Sigma Aldrich (Germany) chemicals.

### 2.2 Synthesis Procedure of RE-doped $V_2O_5$

The pristine, Ho-doped, and Yb-doped $V_2O_5$ were synthesized by facile hydrothermal technique following our previous work [42] . The steps of the whole procedure are illustrated in figure 1. Ammonium metavanadate was dissolved into an equal ratio of de-ionized (DI) water and ethanol to make a 1M concentration. Ho/Yb precursor was added as a doping element into the solution with 1 mol.%, 3 mol.%, and 5 mol.%. Then, the pH level was maintained for making an acidic medium by adding nitrite acid (10wt.%). This solution was stirred for 1 hour at room temperature and transferred to a Teflon-lined hydrothermal autoclave for 24 hours at 100 °C. After the reactions, the products were washed several times with DI water and ethanol using centrifugation and dried for 6 hours in an oven. Finally, the annealing was done at 500 °C for 2 hours to improve the crystallinity of the final product.

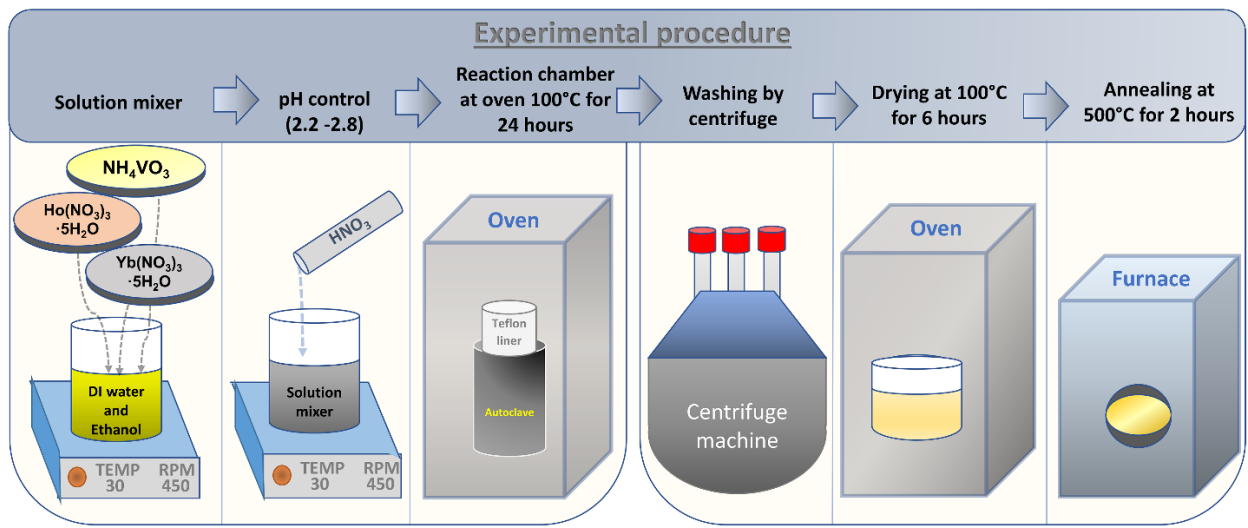

*Figure 1: Experimental procedure of preparing pure and Ho/Yb doped V2O5*

## 2.3 Characterization Techniques

The X-ray diffraction (XRD) technique was used for structural analysis through the Rigaku Smart Lab diffractometer with Cu-Kα radiation (λ=1.5406 Å). Surface morphology was studied by SEM images TESCAN VEGA 3. Qualitative chemical analysis was acquired by Energy dispersive spectroscopy. The UV-Vis absorption spectra were measured in a Lambda 650 Perkin Elmer spectrophotometer in the wavelength range from 200 to 900.

## 2.4 Photocatalytic Performance Test

The photocatalytic performance of pure and Ho/Yb doped $V_2O_5$ was evaluated by adding methylene blue (MB) as a sample of industrial pollutants. A 300 W Xenon lamp was used as the visible light source, and a solar simulator was attached to filter out UV light (λ > 420). We maintained a distance of 20 cm between the light and to sample. The instrumental setup for evaluating photocatalytic activity is illustrated in figure 2. For each experiment, $10^{-4}$ M concentration of MB solution was prepared into 200 ml DI water. A 200 ppm catalyst concentration was added to investigate the effects of pH variation. On the other hand, 500 ppm concentration was used to observe the improvement of photodegradation efficiency in the presence of RE doping in the $V_2O_5$ particles.

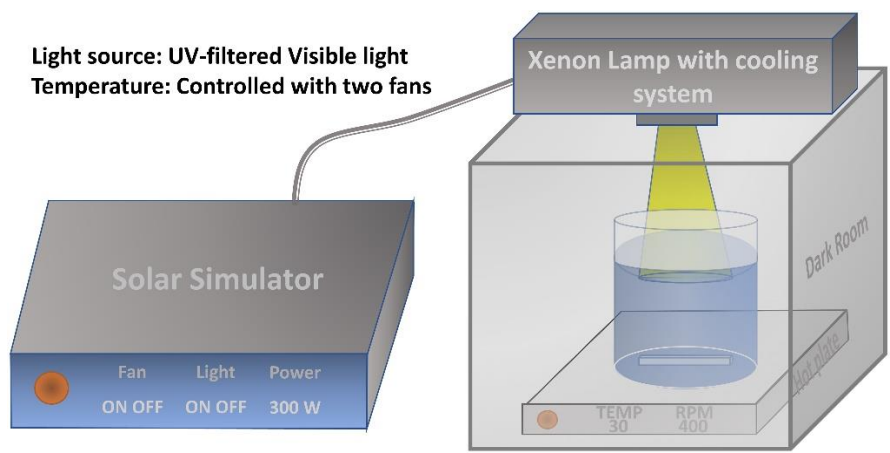

*Figure 2: Equipment arrangement for the photocatalytic performance measurement*

After adding the catalyst, the solution was stirred for 30 min in the dark condition to develop an adsorption-desorption equilibrium between the $V_2O_5$ and MB. After turning the light ON, absorption spectra were measured at 20 minute intervals. For this, a 10 ml solution was separated from the original solution and then centrifuged the liquid at 5000 rpm for 5 mins.

## 3. Result and Discussions

### 3.1 Structural Properties

Figure 3(a) presents the XRD patterns of Ho-doped and Yb-doped $V_2O_5$, along with the pristine condition of $V_2O_5$. It is observed that all peaks of the XRD patterns are matched with standard XRD patterns of JCPDS #41-1426, claiming a stable orthorhombic structure has been formed.[43] No additional peaks are detected in the XRD patterns, meaning that no secondary phases have been identified in the doping samples.

However, the intensity of the peaks of HoVO (Holmium doped $V_2O_5$) and YbVO (Ytterbium doped $V_2O_5$) differ from their pure $V_2O_5$. The intensity of the peaks decreases with increasing doping concentration indicating the lowering of the crystallite size at the expense of the crystallinity of the samples.[35] The difference in the ionic radius between $V^{5+}$ (5.4 nm) and $Ho^{3+}$ (8.9 nm) or $Yb^{3+}$ (8.6 nm) is relatively high, which forms these discrepancies in the crystal structure. It is also observed that the peaks are shifted to lower 2θ values indicating volume expansion occurred in the doped sample compared to pure $V_2O_5$, in Figure 3(b-c). 3HoVO and 3YbVO depict the highest peak shifts with respect to all the samples and prosecute the largest stored internal energy in these two samples; attributed to the influence of RE-doped ions on the V-O bonds, creating a local mismatch in the $V_2O_5$ lattice structure.

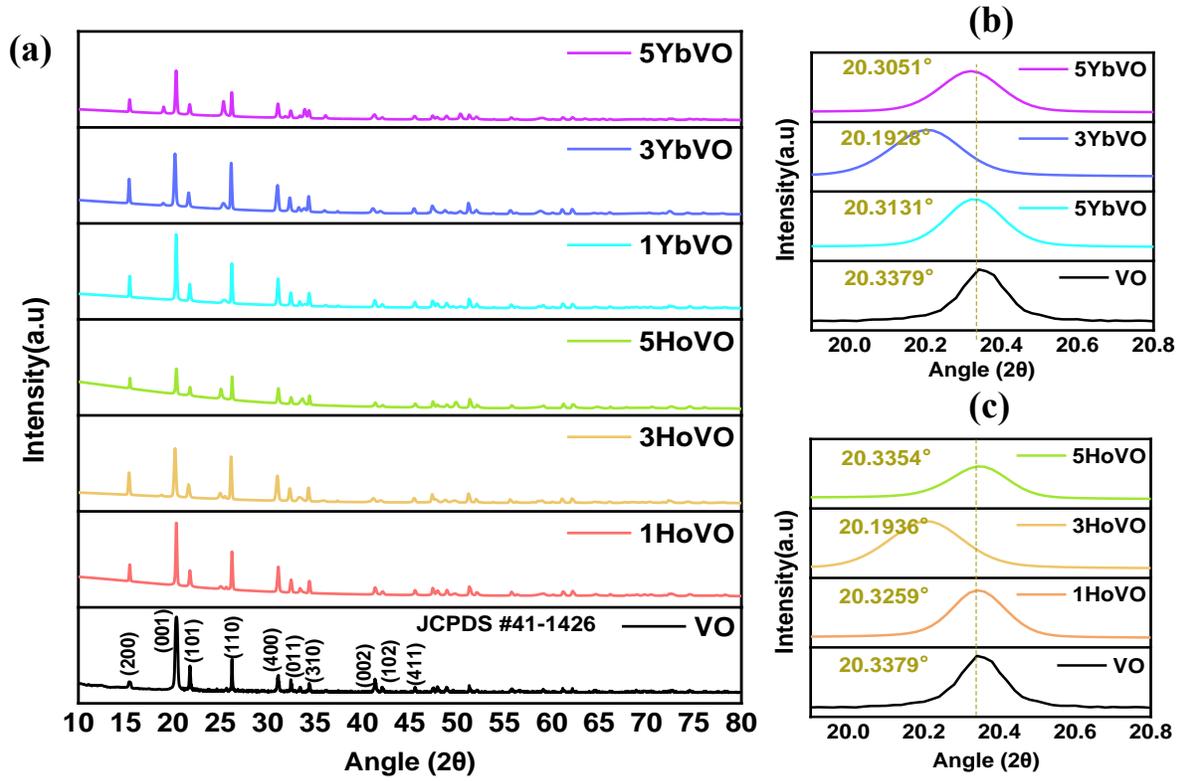

*Figure 3: (a) XRD patterns(a) in the range of 10 -80 degrees, (b-c) magnified pattern in the range of 20-20.8 degrees, of all the doped and undoped $V_2O_5$ samples*

*Table 1: Crystalline parameters of pure and rare-earth doped (Ho/Yb) doped V2O5*

|  | Crystallite size (nm) | | d- Spacing (Å) | | Micro-Strain x 10$^{-3}$ | | Dislocation density x 10$^{-3}$ (nm$^{-2}$) | |
| --- | --- | --- | --- | --- | --- | --- | --- | --- |
| Doping % | **HoVO** | **YbVO** | **HoVO** | **YbVO** | **HoVO** | **YbVO** | **HoVO** | **YbVO** |
| 0 | 71.10 | 71.10 | 4.363 | 4.363 | 2.76 | 2.76 | 1.98 | 1.98 |
| 1 | 52.60 | 48.90 | 4.366 | 4.368 | 3.73 | 4.02 | 3.61 | 4.19 |
| 3 | 36.40 | 38.30 | 4.394 | 4.394 | 5.43 | 5.16 | 7.56 | 6.81 |
| 5 | 48.30 | 45.40 | 4.364 | 4.370 | 4.06 | 4.33 | 4.28 | 4.85 |

The average crystallite size (D) of the compounds was calculated by the Debye-Scherrer formula, given below -

$$D = \frac{0.9\,\lambda}{\beta\,\cos\theta} \qquad (1)$$

Where, $\beta$ is the full-width at half maximum of (001) peak, $\theta$ corresponds to the diffraction angle, and $\lambda$ is the X-rays wavelength of Cu K$\alpha$ (1.5406 Å). Similarly, the d-spacing (d), micro-strain($\varepsilon$), and dislocation density ($\delta$) were also calculated using the following equations and presented in Table (1).

$$d = \frac{n\,\lambda}{2\,\sin\theta} \qquad (2)$$

$$\varepsilon = \frac{\beta}{4\,\tan\theta} \qquad (3)$$

$$\delta = \frac{1}{D^2} \qquad (4)$$

The crystallite size of VO is found to be 71.1 nm, which decreases rapidly with increasing doping percentage for both elements. The lowest crystallite size value is obtained for 3HoVO and 3YbVO, indicating to a decrease in the particle size for 3% doping, which can be explained by the fact that doping ions provide new nucleation sites for particles' formation, which alters the homogeneous nucleation to heterogeneous nucleation. This heterogeneous nucleation is faster than homogeneous nucleation, eventually decreasing the particle size. A similar trend was reported for Cobalt doping in $V_2O_5$.[44] It is reported that the interlayer distance of Zr-doped and Ti-doped $V_2O_5$ nanorods increases due to their large cation size. A similar trend is observed for the present case, the interspacing increases with increasing doping (both Ho and Yb) percentage. The highest spacing estimates for 3mol.% doping concentration of both RE elements in the $V_2O_5$. Moreover, the micro-strain and dislocation density also followed the trends of d-spacing. Both samples are increased with the increment of the doping elements up to 3 mol.%.

### 3.2 Morphological Analysis

The surface morphology of the synthesized pure and Ho/Yb-doped $V_2O_5$ was investigated by taking scanning electron microscopy (SEM) images. Energy dispersive X-ray spectrum (EDX) has also been utilized to confirm the presence of desired elements, such as V, O, Ho, and Yb. In figure 4a, the particle of $V_2O_5$ is in irregular shape where agglomeration occurs due to the high surface energy of these particles. The EDX of $V_2O_5$ is presented in figure 4b, where no additional peaks are observed except V and O. Besides, the atomic percentage of the elements confirmed the successful formation of $V_2O_5$.

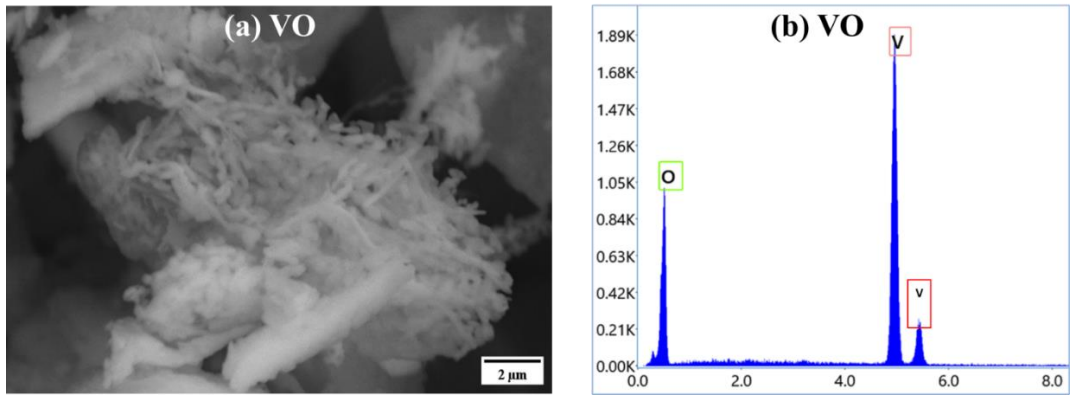

*Figure 4: (a) SEM image (b) EDX spectra of pure $V_2O_5$*

The morphology of the particles has been significantly changed with increasing Ho/Yb ions doping, as illustrated in figure 5. The particle size reduces compared with their pristine V2O5 for both cases. The finding is highly persuasive as doping elements create new nucleation and promote heterogeneous nucleation, which eventually reduces to chance of the growth of the particle. However, the agglomeration is continued for the doped samples.

EDX spectra of Ho/Yb doped $V_2O_5$ are exhibited in figure 6. Although the EDX spectra cannot be conclusive in calculating the exact percentage of elements in the overall sample, it can be an excellent way to examine the presence of expected elements. Our expected elements in these spectra have been clearly visible, and the intensity for Ho and Yb increases with increasing the doping percentage.

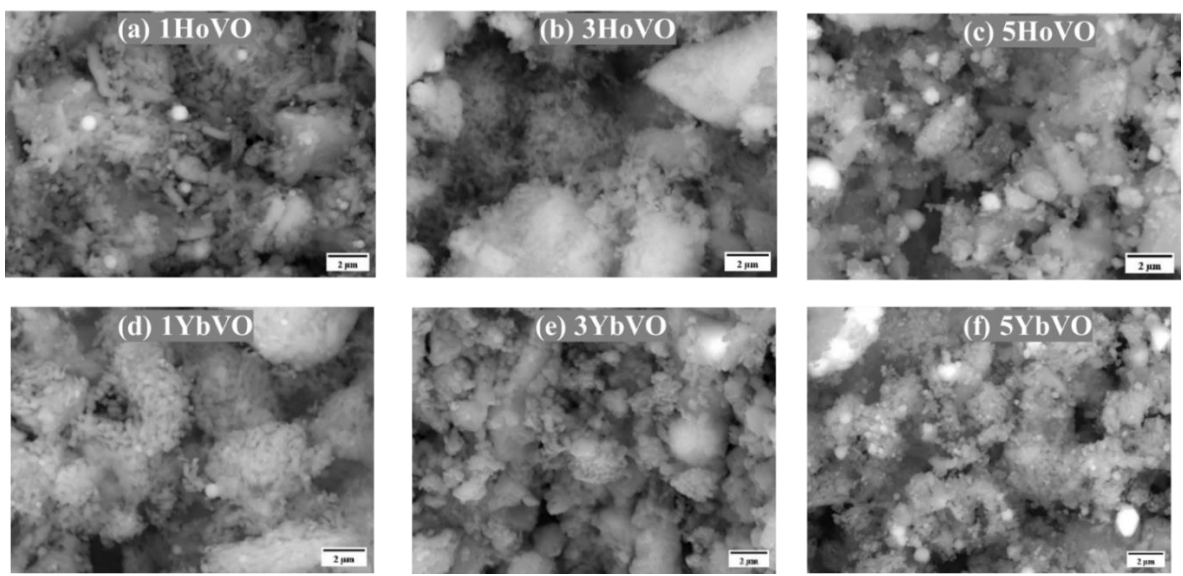

*Figure 5: SEM image of (a) 1mol. % Holmium-doped $V_2O_5$(1HoVO) (b) 3 mol. % Holmium- doped $V_2O_5$ (3HoVO) (c) 5mol. % Holmium-doped $V_2O_5$ (5HoVO) (d) 1mol. % Ytterbium -doped $V_2O_5$ (1YbVO) (e) 3 mol. % Ytterbium -doped $V_2O_5$ (3YbVO) (f) 5mol. % Ytterbium -doped $V_2O_5$ (5YbVO).*

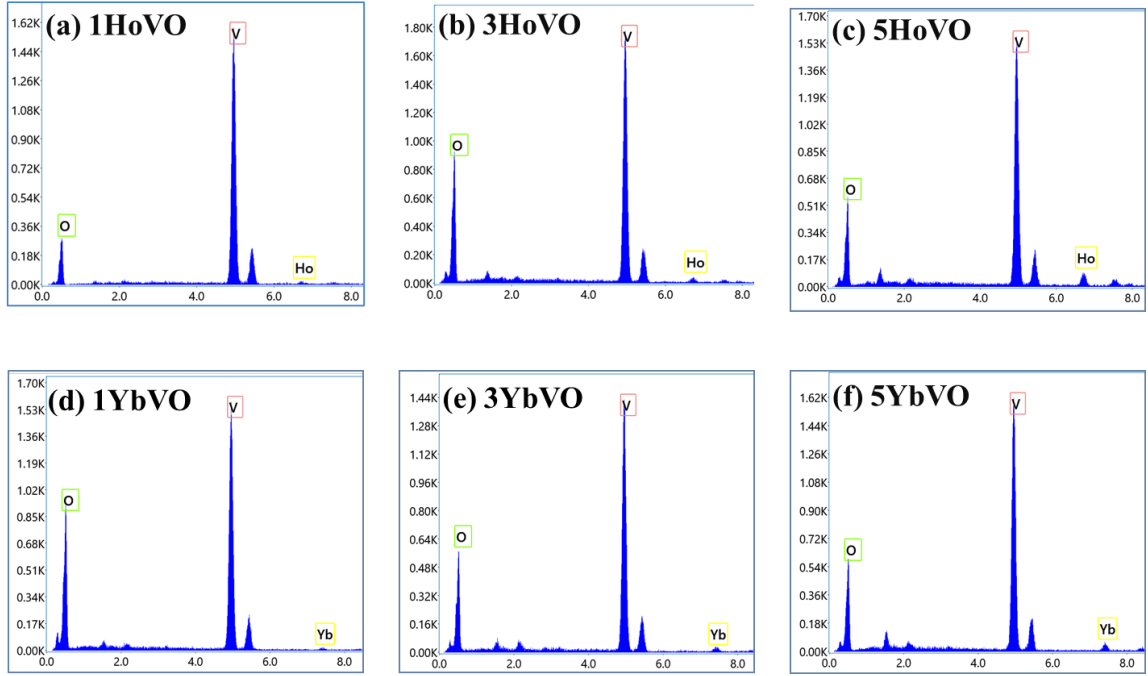

*Figure 6: EDX spectra of (a) 1mol. % Holmium-doped $V_2O_5$(1HoVO) (b) 3 mol. % Holmium- doped $V_2O_5$ (3HoVO) (c) 5mol. % Holmium-doped $V_2O_5$ (5HoVO) (d) 1mol. % Ytterbium -doped $V_2O_5$ (1YbVO) (e) 3 mol. % Ytterbium -doped $V_2O_5$ (3YbVO) (f) 5mol. % Ytterbium -doped $V_2O_5$ (5YbVO).*

### 3.3 Optical Characteristics

UV-visible (UV-vis) absorption spectra of pure, Ho-doped, and Yb-doped $V_2O_5$ are presented in figure 7(a-b). All the samples show strong absorption in the visible range, which is attributed to the electron transition from O (2p) to V (3d).[35,44] For the Holmium/Ytterbium doping, the absorption increases in the UV range due to the unfilled orbitals of RE ions. Besides, the substitution of $V^{5+}$ by $Ho^{3+}$ or $Yb^{3+}$ creates a charge imbalance in the crystal structure, leading to the formation of oxygen vacancies to compensate for the charge imbalance. These defects are also responsible for enhancing the absorption of the doped $V_2O_5$ samples. [43]

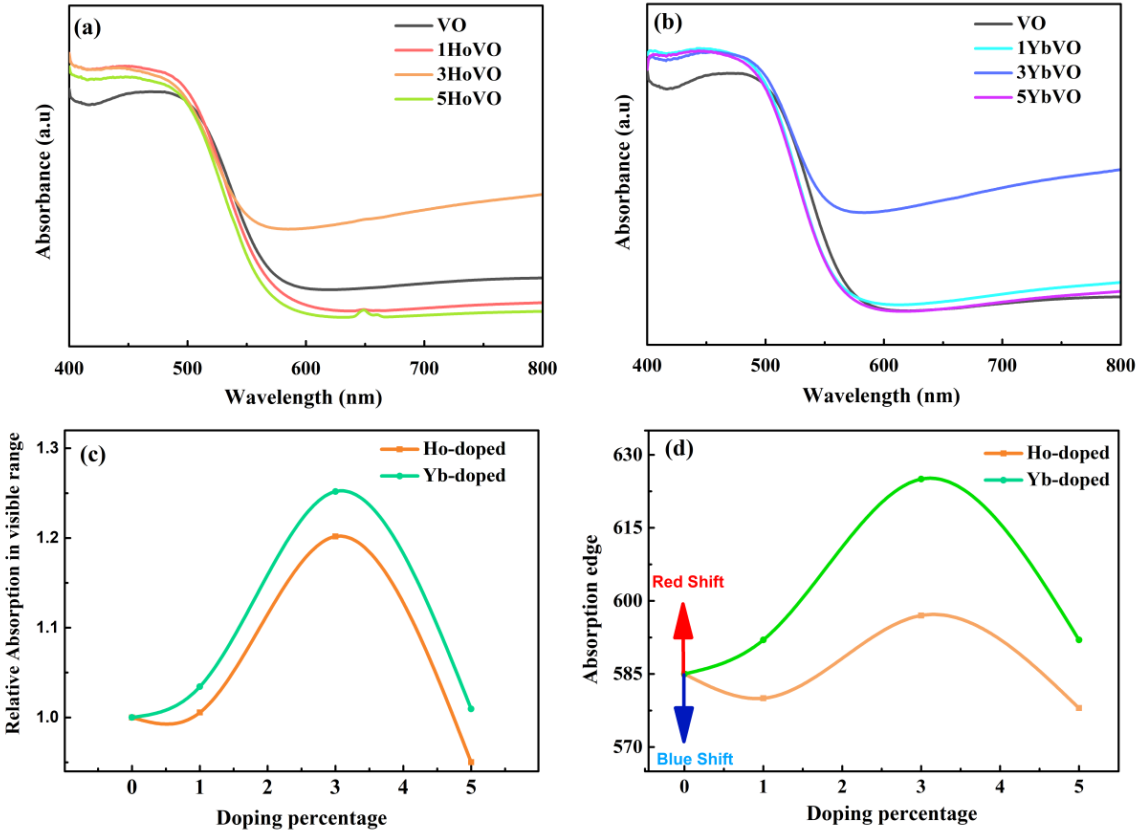

*Figure 7: UV-visible absorption spectra for various (a) Ho-doping and (b) Yb Doping. (c) Calculated relative absorption and (d) absorption edge of Ho-doping and Yb Doping*

The relative absorption (RA) is an essential parameter to observe the improvement of solar efficiency of the prepared sample, as shown in figure 7(c) for the doped samples. It is measured by the following equation in the visible range (380 -750 nm):

$$Relative\ absorption = \frac{Average\ absorption\ of\ desired\ sample\ in\ the\ visible\ range}{Average\ absorption\ of\ pure\ V_2O_5\ sample\ in\ the\ visible\ range} \quad (5)$$

The RA value increases with the increment of doping percentage for the samples up to 3 mol.% and then drops very sharply for 5 mol.% percentage. The 3YVO shows a higher RA value than 3HVO, indicating higher solar efficiency can be achieved in the visible range. This increment of the absorption is highly aligned with the strained conditions due to the lattice distortion we discussed in the structural analysis section 3.1. The absorption edge (AE) is another vital parameter to characterize the optical properties of the samples where absorption discontinuity or absorption limit occurs a sharp discontinuity in the absorption spectrum of a substance, and at that point, the energy of an absorbed photon corresponds to an electronic transition or ionization potential. The AE of the samples has been calculated and illustrated in figure 7(d). Increasing the AE value indicates the 'red shift', and decreasing the value means the 'blue shift'. The YVO samples are shown the 'red shift' for all the doping percentages compared to the parent VO sample, and 3HVO is shown the maximum value of 'red shift'. The finding indicates that Ytterbium-doping expects to be shown better performance than that Holmium-doping.

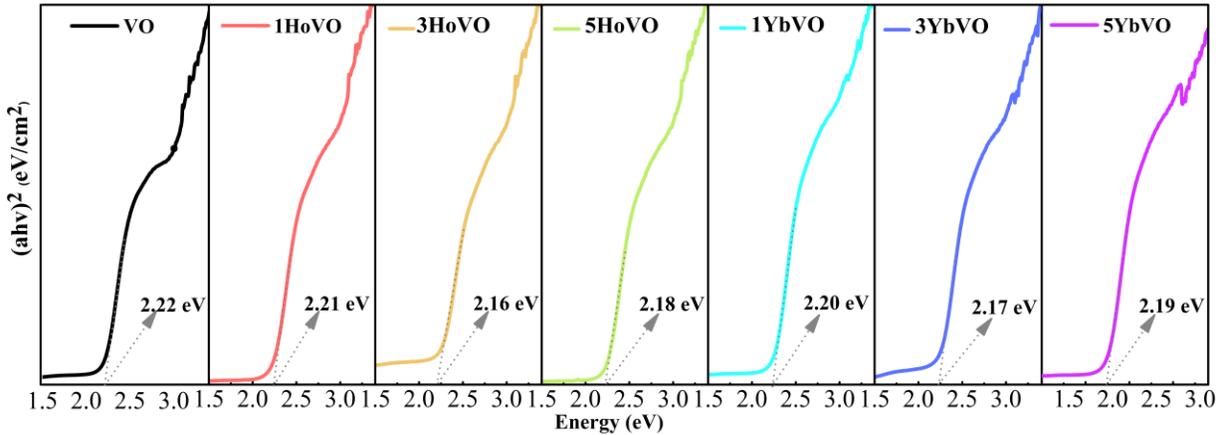

Figure 8: Bandgap measurement of pure, Ho/Yb-doped $V_2O_5$ sample by using Tauc Formula

The bandgap of the pure and doped samples has been measured using the Tauc relationship,

$$\alpha h\vartheta = A\,(h\vartheta - E_g)^n \qquad (6)$$

where α is the absorption coefficient, hν defines the photon energy, *A* suggests a proportional constant, and $E_g$ represents the bandgap of the sample. Two different allowed transitions are classified by using two different 'n' values: n = 2 for direct and n = ½ for indirect bandgap. The calculated bandgap of the samples using the Tauc plot is depicted in figure 8. It is observed that the bandgap reduces for all the doping samples compared to their parent VO. The reduction is caused by introducing defects in the crystal system of pure VO by doping the foreign elements. These foreign elements form additional energy states within the bandgap that create intermediate paths for electron transitions and delay the electron recombination rate.[45] The maximum decrements of the bandgap are found to be 2.16 eV (3HoVO) and 2.17 eV (3YbVO), which is consistent with our discussion regarding the highest value of RA and AE.

### 3.4 Photocatalytic Measurements

Methylene blue (MB) is a symbol of pollutants for measuring photocatalysis efficiency. It shows two characteristic peaks in the visible absorption spectra at 613 nm and 664 nm. The peak at 664 nm was considered to evaluate the degradation of the MB. As the decrement of the intensity is proportional to the degradation of the product, the efficiency of photo degradation is calculated by using the following equation:

$$Dye\ removal\ efficiency = \left(1 - \frac{C_t}{C_o}\right) \times 100 \qquad (7)$$

Here, $C_0$ represents the initial concentration and $C_t$ defines the concentration of the sampling time. Langmuir Hinshelwood (L-H) kinetics can be used to measure the kinetics of our synthesized catalysts[46], given below:

$$r = -\frac{dC}{dt} = \frac{k_T K C_t}{1 + K C_t} \qquad (8)$$

Where, $K_r$ defines time-dependent reaction rate, K is the equilibrium constant for adsorption. We have used a very low concentration of MB ($10^{-4}$ M), the equation (8) can be approximated to the first-order kinetic reaction[47,48], presented in equation (9):

$$\ln\left(\frac{C_0}{C_t}\right) = kt \qquad (9)$$

Here, k defines pseudo-first order rate constant. This rate constant is calculated from the slope of the $\ln\left(\frac{C_0}{C_t}\right)$ vs time curve.

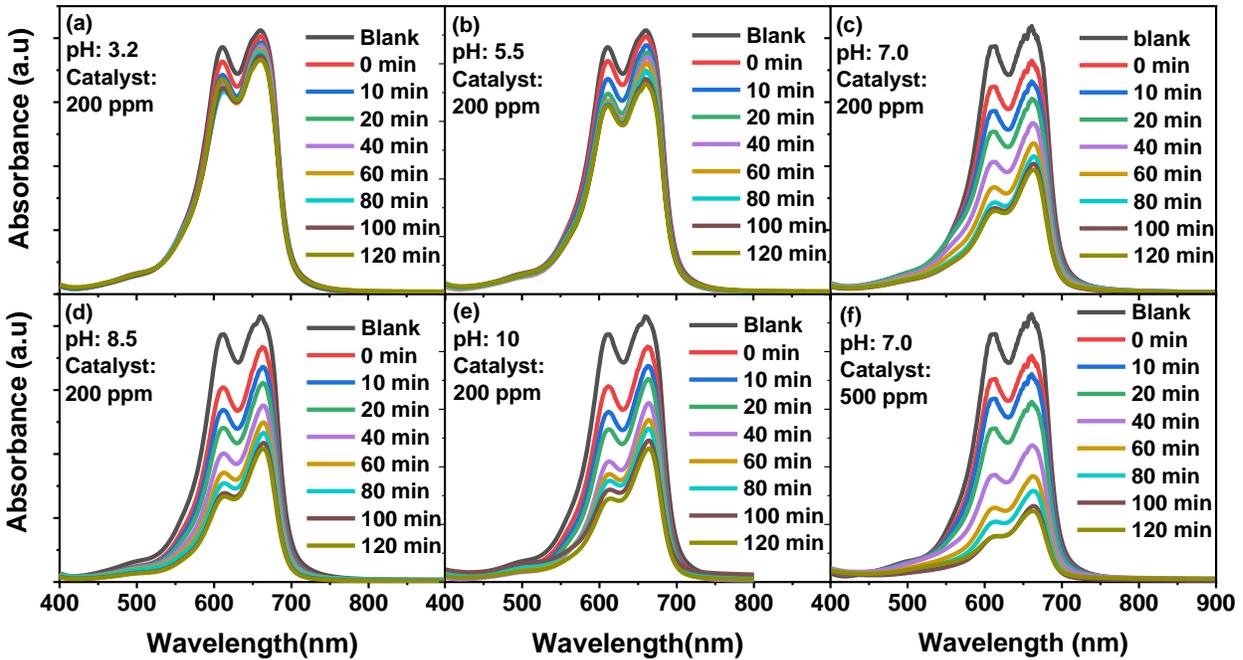

*Figure 9: Absorption spectra of MB degradation under various pH levels and catalyst concentrations: (a) as received, (b) pH = 5.5, (c) pH = 7.0, (d) pH = 8.5, (e) pH = 10 and (f) pH = 7 and catalyst = 500 ppm*

### 3.4.1 Effects of pH

Industrial textile dyes contain various contaminants having a wide range of pH levels. So, it is necessary to check the performance of $V_2O_5$ on the basis of different pH levels. The prepared sample showed a very low pH value (3.3) due to its synthesis process, wherein a pH of 2.5 was maintained. We examine the photodegradation efficiency of $V_2O_5$ as a function of a wide range of pH levels of 3.3 – 10.0. We set a 200 ppm concentration for the catalyst. The wavelength dependence of absorptions for various pH values is represented in Figure 9.

Time-dependent degradation of the MB in the presence of various pH levels is illustrated in Figure 10(a). The degradation is very low at low pH values (acidic medium), whereas the efficiency improves exponentially from 11% to 52% with a rising pH level from 5.5 to 7.0. The efficiency decreases afterward in the alkaline medium. The Rate constants were calculated for different pH levels, as shown in Figure 10(b). The rate constant curves clearly indicate that the highest degradation is found for pH of 7.0, indicating that the neutral condition or slightly alkaline medium is good for getting better photodegradation efficiency of $V_2O_5$. The overall degradation efficiency and rate constant values are plotted in Figure 11.

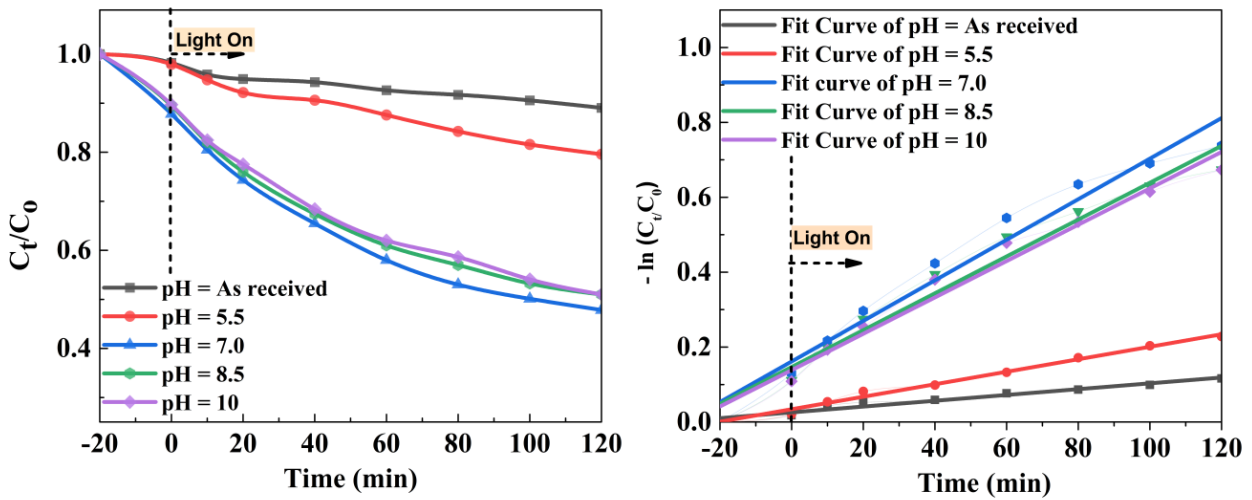

*Figure 10: (a) Photocatalytic degradation efficiency and (b) degradation rate constant of MB by V2O5 under various pH levels*

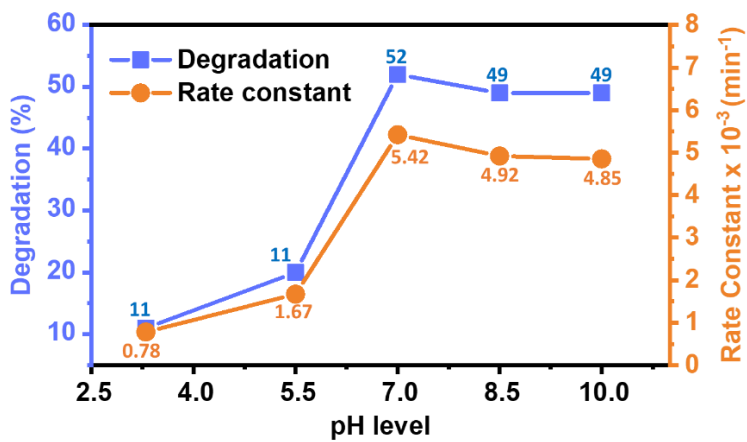

*Figure 11: Degradation efficiency (blue) and Rate constants (orange) of $V_2O_5$ regarding to various pH level*

This phenomenon can be explained by electrostatic interactions between the catalysts and pollutants, as illustrated in figure 12. One of the important parameters to observe the pH effect is the point zero charge (PZC) of the catalysts, where the net charge of the particle surface is zero for a certain pH level.[49] The surface of the particle will be positively charged below this pH level and negatively charged above the pH level. As the MB is cationic, the surface of the catalyst must be negatively charged so that the catalyst can come in contact with the pollutants. At an acidic medium (pH < $pH_{PZC}$), the surface of $V_2O_5$ is positively charged, repels the MB, and cannot degrade. However, in the neutral or slightly alkaline medium, the surface of catalysts is negatively charged, so adsorption can occur. In our case, the degradation rate was changed dramatically by changing the pH value from 5.5 to 7. So, one can conclude that the PZC of $V_2O_5$ lies in between pH of 5.5 – 7. It should be noted that the photodegradation efficiency for a very high pH level (11-14) is not considered in this study. However, many reports claimed that the samples show low efficiency at this extreme pH level.[24,49,50] This may be attributed to the neutralization of cationic pollutants due to high OH concentration and restriction to contact with the catalyst.

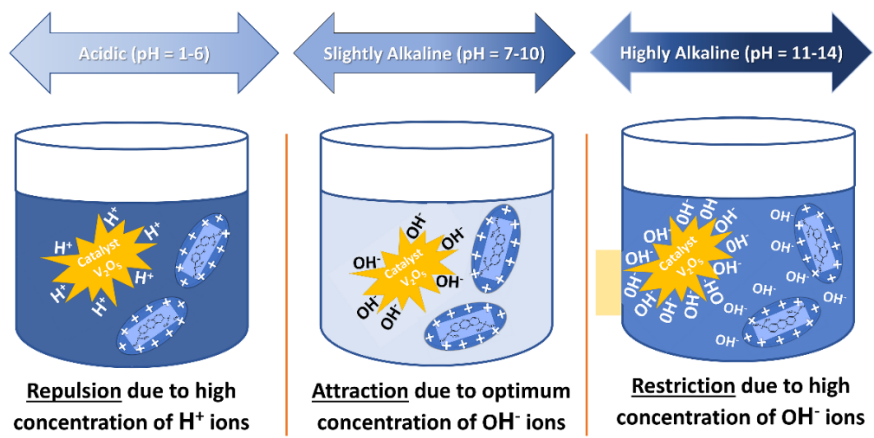

*Figure 12: Mechanism of pH effects on photodegradation efficiency of $V_2O_5$*

### 3.4.2 Effects of Ho-doping and Yb-doping

The photocatalytic degradation efficiency of Ho-doped and Yb-doped $V_2O_5$ have been measured using the intensity of the characteristic peak of MB in the UV-visible absorption spectra, as presented in Figure 13. A catalyst concentration of 500 ppm is used to observe the efficiency of the samples.

Figure 14(a-b) exhibits the degradation efficiencies and rate constants of 1HVO and 3HVO as a function of time. After 2 hours of visible light irradiation, the degradation increases while it is decreased for 5HVO producing the highest degradation for 3HVO among all samples. The rate constants are found to be 0.012 min$^{-1}$, 0.020 min$^{-1}$, and 0.008 min$^{-1}$ for 1HVO, 3HVO, and 5HVO, respectively. Interestingly, the rate constant of 3HVO is almost double compared to its pristine V2O5 sample. The degradation percentages are estimated at 78, 93, and 68% for 1HVO, 3HVO, and 5HVO, respectively. Similar trends are observed for Yb doped V2O5 crystal structure as depicted in figure 14(c-d). The estimated degradation percentages are 81, 95, and 77% for 1YbVO, 3YbVO, and 5YbVO, respectively. The rate constant is also improved with increasing Yb percentage up to 3 mol.% and then reduced for 5YbVO. Thus, 3HoVO and 3YbVO can be considered promising candidates for photocatalysis.

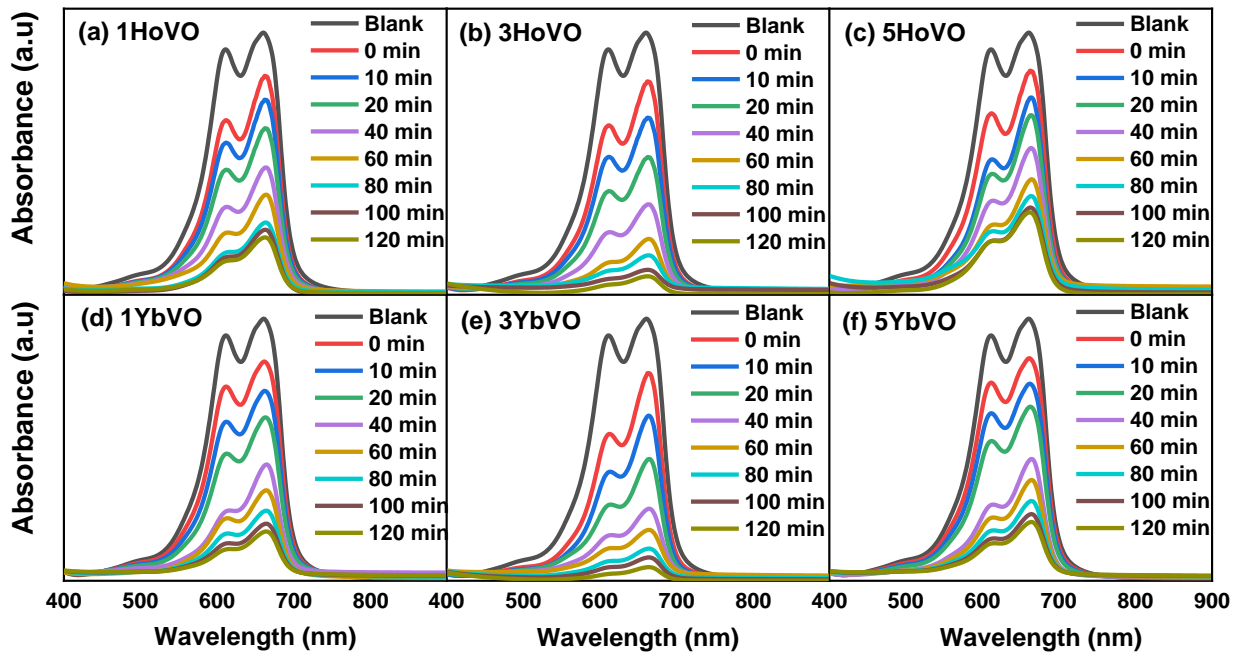

*Figure 13: UV-vis absorption spectra of (a) 1mol.% Ho (b) 3mol.% Ho (c) 5 mol.% Ho (d) 1mol.% Yb (e) 3mol.% Yb (c) 5 mol.% Yb-doped $V_2O_5$*

A comparison chart of degradation efficiency and the rate constant is illustrated in figure 15 for better understanding. It is shown that 3 mol.% exhibits the best performance in degrading the pollutants. As the ionic size differences of RE and vanadium are substantial, the crystallinity decreases for high doping percentages such as 5HoVO and 5YbVO. This damaged crystallinity reduces the light absorption in the visible range (Figure 7), consequently reducing the degradation efficiency.

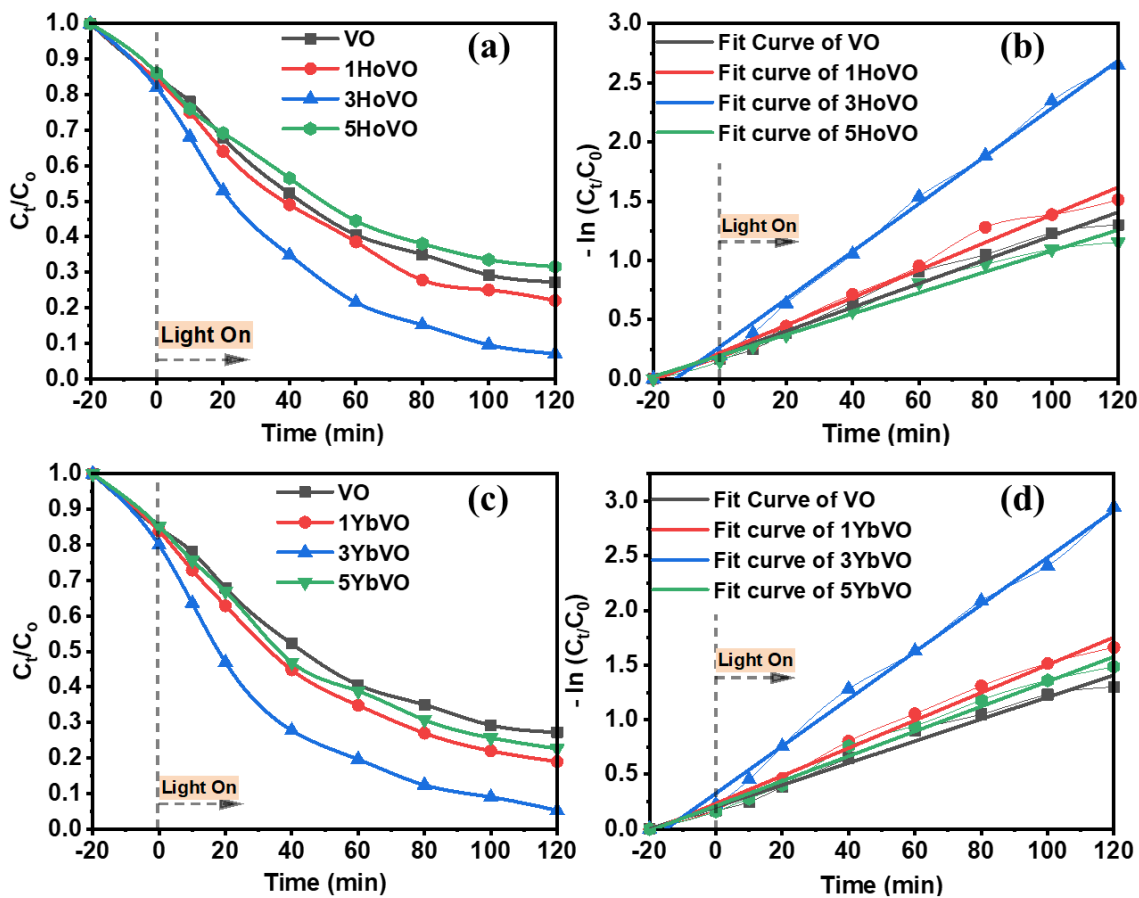

*Figure 14: Degradation of MB as a function of time for (a) Ho-doped (c) Yb-doped $V_2O_5$ samples. Degradation rate constants as a function of time for (a) Ho-doped (c) Yb-doped $V_2O_5$ samples.*

We have compared our results with available reported RE elements doped $V_2O_5$, as depicted in Table 2. Various measuring factors affect the degradation efficiency and rate constant such as catalyst concentration, pollutants concentration, degradation time, and energy source. A high concentration of catalyst and a low concentration of pollutants exhibits a high value of degradation percentage.[33] Instead of using a higher MB concentration, we obtain the competitive degradation performance among all the $V_2O_5$ samples.

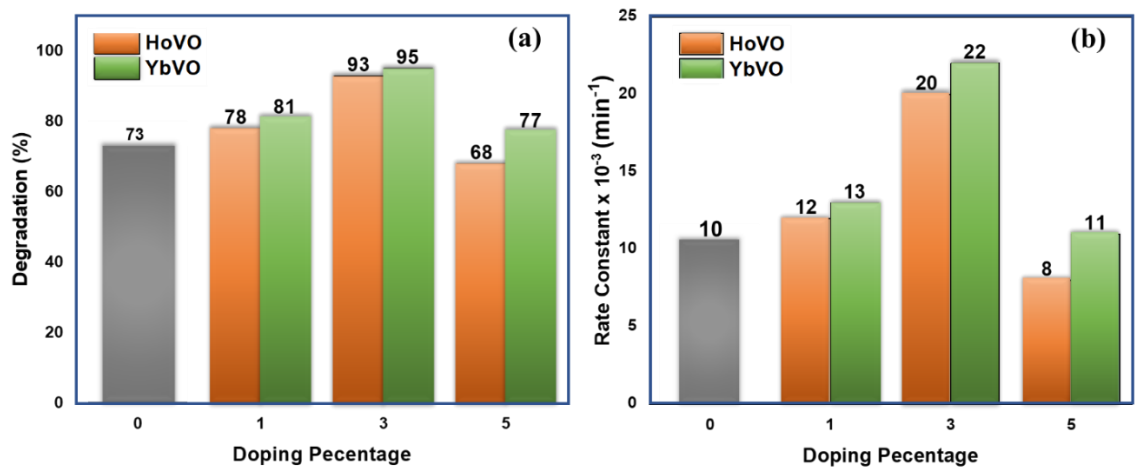

*Figure 15: (a) Degradation efficiency (b) degradation rate constant Comparison of pure, Ho and Yb doped V$_2$O$_5$.*

Although Nd$^{3+}$-doping showed 99% degradation, the concentration of catalyst was double, the concentration of MB was one order lowers, and the light source was in the UV range. However, our obtained degradation rate constants of 3HVO and 3YbVO are the highest compared to all listed V$_2$O$_5$ particles. It is evident that 3HoVO and 3YbVO are the best candidates for photodegradation in V$_2$O$_5$.

*Table 2: An overview of the effects of various ions on V2O5 for the photodegradation efficiency of MB*

| V$_2$O$_5$ Catalysts with various doping | Catalysts Conc. (ppm) | MB Conc. (Molarity) | Light Source | Degradation efficiency (%) | Rate constant (min$^{-1}$) x 10$^{-3}$ | Ref |
|---|---|---|---|---|---|---|
| 5 wt.% Gd | 100 | 1.6 x 10$^{-5}$ | Visible light | 46 | 5.50 | [35] |
| 5 wt.% Ti | 200 | 1.0 x 10$^{-5}$ | Visible light | 82 | 14.0 | [43] |
| 3 mol.% Sn | 200 | - | Visible light | 95 | 15.3 | [28] |
| 5 wt.% Co | 500 | 1.0 x 10$^{-4}$ | Visible light | 91 | 20.0 | [44] |
| 7 mol.% Nd | 1000 | 1.6 x 10$^{-5}$ | UV light | 99 | 14.0 | [33] |
| 3 mol.% Ho | 500 | 1.0 x 10$^{-4}$ | Visible light | 93 | 20.0 | This |
| 3 mol.% Yb | 500 | 1.0 x 10$^{-4}$ | Visible light | 95 | 22.0 | This |

### 3.5 Plausible Mechanism of degradation

The possible mechanism of degradation of pollutants using the catalyst and dye MB is illustrated in figure 16. The photons eject electrons from the valence band (VB) of the catalyst surface and move to the conduction band (CV). The recombination occurs between the electron and the hole itself or by the surface charges. Precisely controlling the recombination rate measures the performance of photocatalysis.

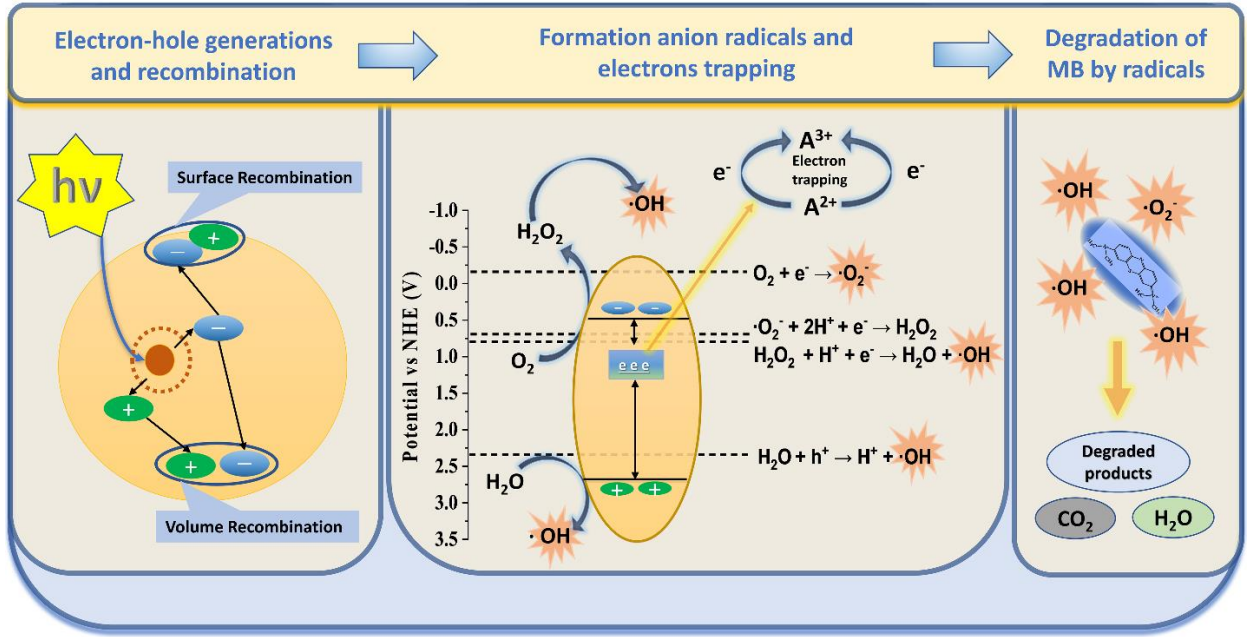

*Figure 16: Schematic diagram demonstrating how rare-earth doped $V_2O_5$ degrades the pollutant efficiently*

The potential of the conduction band ($E_{CB}$) and valance band ($E_{VB}$) for the $V_2O_5$ can be calculated by the following equations[50],

$$E_{CB} = \chi + E_C - 0.5\, E_g$$

$$E_{VB} = E_g + E_{CB}$$

Here, $\chi$ is the electronegativity of the $V_2O_5$ (6.1 eV)[51], $E_g$ defines the bandgap of the $V_2O_5$ (2.22 eV), and $E_C$ represents the energy of free electrons in the hydrogen scale (4.5 eV). The data provides the value of 0.49 and 2.71 eV for $E_{CB}$ and $E_{VB}$, respectively. Since the addition of doping elements did not change the bandgap in a wide range, the $E_{CB}$ and $E_{CB}$ will remain almost the same as the undoped sample. However, these doping elements create various defects by forming interstitial oxygen and vacancies of vanadium, which introduce additional energy states within the bandgap. These energy states are considered at trapping sites to suppress the electron-hole recombination and lengthen the lifetime of photo-excitons.

Apart from the bandgap reduction and defect formation phenomena, the RE ions form complex compounds with the organic dye that also perpetuate the pollutant's degradation reaction. It is attributed to having partially filled 4f orbital in the electronic configuration of the RE elements. Moreover, the work function of $V_2O_5$ decreases with increasing RE doping.[33] This decrement suggests that the Fermi level of $V_2O_5$ moves towards the CB and makes electron move easier to the acceptors. Finally, these photo-generated electrons can capture oxygen ($O_2$) and form superoxide radicals ($^•O_2^−$), which would continue further reactions to reduce the recombination rate. On the other hand, holes in the valance band can react with water ($H_2O$) to form hydroxyl radical ($^•OH$). These highly reactive radicals attack the pollutants to degrade to generate $CO_2$ and $H_2O$. The degradation pathway of MB in the presence of free radicals is thoroughly discussed in the previous report. [48] Thus, the degradation of pollutants has been enhanced in the presence of RE elements.

## 4. Conclusions

The Ho and Yb doped $V_2O_5$ have been produced using an environment-friendly facile mild hydrothermal method, and the effect of doping elements on the photocatalytic activity of $V_2O_5$ has been studied in comparison with pristine $V_2O_5$. The doping of RE ions raises the absorption in the UV range due to their unfilled orbitals and customizes oxygen vacancies by creating a charge imbalance in the crystal to reimburse the charge imbalance, enhancing the absorption. The highest relative absorption (RA) and absorption edge (AD) are estimated at 3 mol.% Yb doped $V_2O_5$ sample, consequently the highest degradation efficiency and rate constants are also determined for 3YbVO of 95% and 22, respectively. The bandgap of doped samples decreases with increasing percentages of dopant, and the highest changes are determined at 2.16 eV (3HoVO) and 2.17 eV (3YbVO), which is approved by the higher RA and AD values estimated for 3YbVO. It is noteworthy that the significant effect of pH on degradation has been studied in detail. The improvement of degradation efficiency estimates exponentially from 11% to 52% by varying the pH level from 5.5 to 7.0. The findings revealed that the maximum efficiency can be achieved for a pH of 7.0, which indicates that the neutral condition or slightly alkaline medium is good for getting better photodegradation efficiency of $V_2O_5$. The photocatalytic study in detail with doped RE ions has been uncovered, and the highest degradation efficiency is found to be 93% and 95% for the 3 mol.% of Ho and 3 mol.% of Yb-doped $V_2O_5$ samples, respectively, within 2 hours. The study suggests that 3YbVO and 3HoVO can be considered very efficient catalysts for mitigating serious environmental pollutant issues.

**Conflicts of interest**

The authors are declared no conflicts of interest.

**Author contributions statement:**

**M. H. Kabir**: Conceptualization, Methodology, Investigation, Software, Data Analysis, Writing-Original draft preparation, Reviewing and Editing. **M. Z. Hossain**: Methodology, Photocatalytic experiment, **M. A. Jalil**: Methodology, Software, Photocatalytic experiment, Editing and Reviewing, **M.M. Hossain**: Reviewing and Editing, **M. A. Ali:** Reviewing and Editing, **M. U. Khandaker:** Reviewing and Editing, **D. Jana**: Reviewing and Editing, **M. M. Rahman:** SEM image and reviewing, **M. K. Hossain:** SEM image and reviewing, **M.M. Uddin**: Supervision, Conceptualization, Original draft preparation, Reviewing and Editing,

**Acknowledgement**

The authors are grateful to the Directorate of Research and Extension (DRE), Chittagong University of Engineering and Technology (CUET), Chattogram-4349, Bangladesh and University of Grants Commission (UGC) for arranging financial assistance under grant numbers CUET DRE (CUET/DRE/2018-19/PHY/008) and 37.01.0000.073.07.017.22.209 for arranging the financial support for this work. We are also thankful for the laboratory support of the Materials Science Division, Atomic Energy Centre, Dhaka 1000, Bangladesh for the experimental support.